\documentclass[journal=nalefd,manuscript=letter]{achemso}
\usepackage[version=3]{mhchem}
\usepackage{amsmath}
\usepackage{amsfonts}
\usepackage{amssymb}
\usepackage{graphicx}
\usepackage{color}

\title[] {InAs nanowire transistors with multiple, independent wrap-gate segments}

\author{A.M. Burke}
\affiliation{School of Physics, University of New South Wales, Sydney NSW 2052, Australia}
\alsoaffiliation{Solid State Physics/Nanometer Structure Consortium (nmC@LU), Lund University SE-221 00 Lund, Sweden}
\author{D.J. Carrad}
\author{J.G. Gluschke}
\affiliation{School of Physics, University of New South Wales, Sydney NSW 2052, Australia}
\author{K. Storm}
\author{S. Fahlvik Svensson}
\author{H. Linke}
\author{L. Samuelson}
\affiliation{Solid State Physics/Nanometer Structure Consortium (nmC@LU), Lund University SE-221 00 Lund, Sweden}
\author{A.P. Micolich}
\email{adam.micolich@nanoelectronics.physics.unsw.edu.au}
\affiliation{School of Physics, University of New South Wales, Sydney NSW 2052, Australia}
\date{\today}

\begin{document}

\begin{abstract}

We report a method for making horizontal wrap-gate nanowire transistors with up to four independently controllable wrap-gated segments. While the step up to two independent wrap-gates requires a major change in fabrication methodology, a key advantage to this new approach, and the horizontal orientation more generally, is that achieving more than two wrap-gate segments then requires no extra fabrication steps. This is in contrast to the vertical orientation, where a significant subset of the fabrication steps needs to be repeated for each additional gate. We show that cross-talk between adjacent wrap-gate segments is negligible despite separations less than $200$~nm. We also demonstrate the ability to make multiple wrap-gate transistors on a single nanowire using the exact same process. The excellent scalability potential of horizontal wrap-gate nanowire transistors makes them highly favourable for the development of advanced nanowire devices and possible integration with vertical wrap-gate nanowire transistors in 3D nanowire network architectures.

\end{abstract}

\maketitle

A driving force in electronics is the miniaturization of the field-effect transistor, a device where the current flowing through a semiconductor channel is electrostatically controlled by the voltage applied to a metal gate electrode. The traditional planar Metal-Oxide-Semiconductor Field-Effect Transistor (MOSFET) features a two-dimensional channel separated from the gate by a thin insulating oxide in a parallel-plate capacitor arrangement. With the reduction in gate length below $50$~nm, the electrostatics of gate-channel coupling in these planar structures severely compromises electrical performance.\cite{FerainNat11} This has fueled a push towards more advanced transistor designs, e.g., Fin-FETs and trigate FETs, where the gate is `folded' around the channel to enhance the coupling, mitigate short channel effects, and improve performance and scalability.\cite{FerainNat11, ChauNatMat07} Such structures are now used in Intel's $22$~nm technology node.

From an electrostatic perspective, the ultimate configuration involves a gate wrapped around the entire channel.\cite{ParkTED02, KhanalNL07} These structures can be made via traditional top-down approaches using silicon-on-insulator wafers\cite{ColingeEDM90, LeobandungJVSTB97, SinghEDL06} but involve advanced processing strategies. Self-assembled nanowires grown vertically from a substrate by chemical vapor deposition\cite{HuACR99, ThelanderMT06} present an interesting alternative. Their geometry is particularly favorable for making concentric `wrap-gates',\cite{NgNL04, GoldbergerNL06} a feature that has been exploited to develop high-performance nanowire transistor arrays\cite{BryllertEDL06, WernerssonPIEEE10} for potential industrial applications. Horizontally-oriented wrap-gate nanowire transistors have also been developed using conventional metal/oxide gate formulations\cite{StormNL12, DharaAPL11, ZhangNL06}, epitaxially grown gates\cite{LauhonNat02} and polymer electrolyte gates.\cite{LiangNL12, CarradNL14}

In forging ahead towards goals of both more complex nanowire devices\cite{AppenzellerTED08} and architectures involving 3D nanowire transistor networks\cite{FerrySci08, WernerssonPIEEE10} the next logical step is to develop methods for making wrap-gate nanowire structures featuring multiple, independently-controllable wrap-gate segments. We report a method for making a horizontal wrap-gate nanowire transistor with up to four independently controllable wrap-gate segments and a single horizontal nanowire featuring two independent wrap-gated transistors extending from a common source/drain contact. In this scheme, we limit the technology to fabricating discrete devices with multiple wrap-gates, while the more complicated vertical processing more easily lends itself to processing and fabrication of arrays of identical nanowire devices or circuits. We show a key advantage of the horizontal orientation: scalability. The step up to two independent wrap-gates requires a substantial change on earlier methods,\cite{StormNL12} but once this change is made, devices with more than two wrap-gates can be made without incurring any additional processing steps. This is in stark contrast to the vertical orientation, where each additional gate entails a repetition of a significant subset of the process steps. As a result, while vertical wrap-gate nanowire transistors with two independent gate segments have been developed,\cite{LiEDL11} further scaling in this orientation is likely to be challenging. Returning our focus to 3D architectures, this naturally puts the onus back on the horizontal orientation to carry the scalability burden, hence the importance of research in this direction. Additionally, there is significant scope for the development of both conventional\cite{HuangSci01, YanNat11} and novel\cite{EndoEDL09, BeiuTNN03} logic circuits using multiple wrap-gate nanowire transistors.

\begin{figure}
\includegraphics[width=14cm]{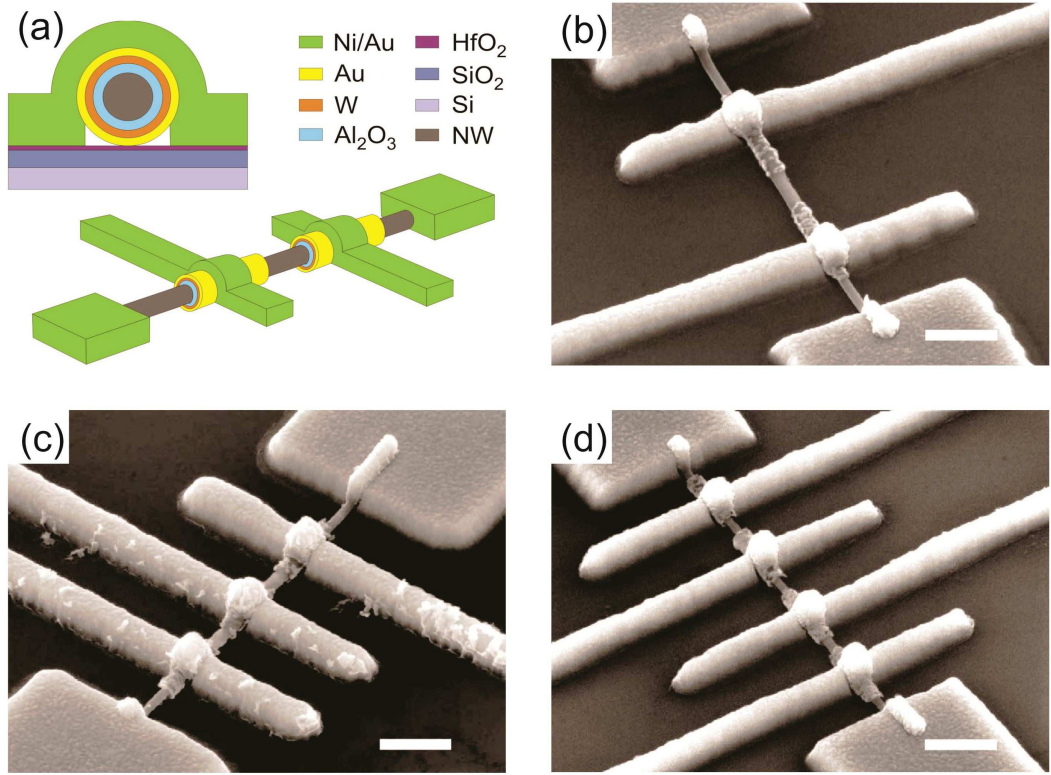}
\caption{(a) Schematic of a horizontal wrap-gate nanowire transistor with two independently-controllable wrap-gate segments. The nanowire (grey) is interfaced by a pair of source and drain electrodes (yellow) and wrap-gate interconnects (green). The cross-section is presented in the inset. (b/c/d) Scanning electron micrographs of nanowire transistors with (b) two, (c) three and (d) four independent wrap-gate segments, respectively. The images shown are from the actual devices studied here, obtained after measurements were completed. The device in (b) was deliberately chosen to highlight that poor wrap-gate appearance does not necessarily correlate with poor electrical performance -- see text. The scale bars in each panel are $500$~nm long.}
\end{figure}

Figure~1(b-d) show scanning electron micrographs of our two-, three- and four-gate nanowire wrap-gate transistors. A corresponding schematic for the device architecture is shown in Fig.~1(a). The fabrication process was inspired by previous work by Storm {\it et al.},\cite{StormNL12} but it is vital to note that it is impossible to make multiple gates using a single EBL resist approach.\cite{StormNL12} This is because each area of the wrap-gate that is etched away necessarily results in a metallic contact to the nanowire after metal deposition. As a result, a number of significant process variations are required to enable the production of multiple wrap-gate segments; amongst them is the need to use multiple EBL resists to avoid device shorting, changes to outer oxide removal and a very different approach to setting wrap-gate segment length. The first fabrication step for the deposited nanowires is a $30$~s etch in buffered hydrofluoric acid (BHF) to remove the entire outer \ce{Al2O3} layer, which is no longer required after deposition. The underlaying substrate is protected during this etch by the \ce{HfO2} layer.\cite{StormNL12} Electron-beam lithography (EBL) was performed using a Raith-150 two system in two separate stages; these two stages are required irrespective of the number of wrap-gate segments or electrical contacts required. In the first stage we expose only the segments of the wrap-gate that need removal. This includes the gaps between wrap-gate segments and the points where source or drain contacts meet the nanowire. The clean gaps between gates show that the outer oxide and Au/W wrap-gate layers have been completely removed. The effective removal of the Au/W is further confirmed by the absence of any current leakage between adjacent wrap-gate segments; the removal of the outer oxide is therefore also confirmed as its removal is pre-requisite to etching the wrap-gate metallization. A $600$~nm layer of polymethylmethacrylate(PMMA) EBL resist was used for the second layer to define contacts to the nanowire source, drain and gates. The thick resist layer is needed to ensure continuity of the deposited metal across the wrap-gate segments. Full details of the device fabrication are given in the Supporting Information.

Electrical characterization was performed at a temperature $T~=~77$~K to reduce drift and hysteresis due to charge trapping at the \ce{Al2O3}/InAs interface. The source-drain current $I_{sd}$ was measured using low frequency a.c. lock-in techniques with an excitation voltage $V_{sd} = 30$~mV at $73$~Hz. All gates were initially tested using a Keithley K2400 voltage source to ensure the leakage current remained at an acceptable level ($< ~100$~pA) over their entire working range. The SRS830 digital to analog converter (DAC) voltage outputs were used thereafter to supply either the voltage $V_{g,n}$ to the $n$~th wrap-gate segment or the voltage $V_{bg}$ to the n$^{+}$-Si back-gate. Low pass RCR filters and ground isolating circuits were used on each DAC output to prevent ground loops and minimize gate noise in the measurements. Two SRS830 lock-ins were used to simultaneously measure each respective transistor's drain current for the common-source nanowire transistor pair.

\begin{figure}
\includegraphics[width=8cm] {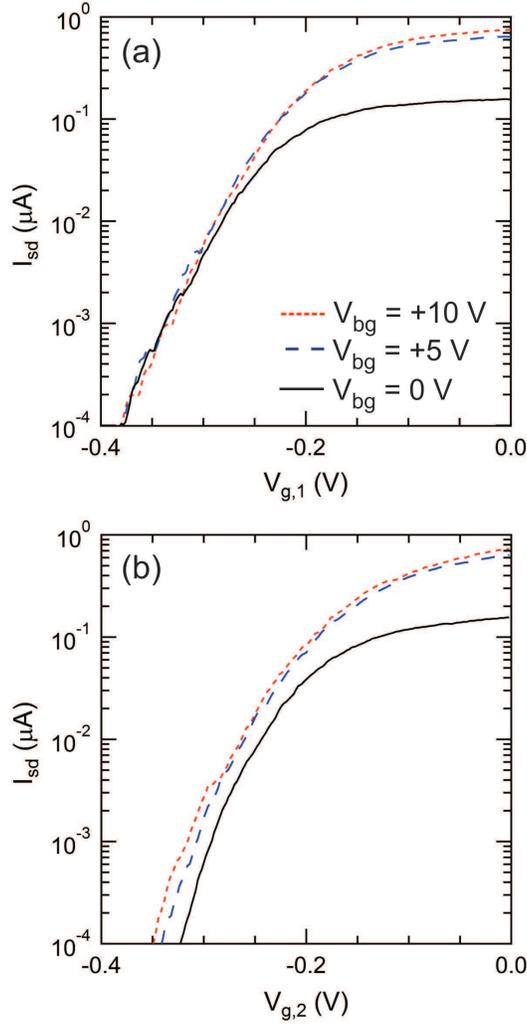}
\caption{Source-drain current $I_{sd}$ vs wrap-gate voltage: (a) $V_{g,1}$ on Gate~1, and (b) $V_{g,2}$ on Gate~2 for the device in Fig.~1(b) for back-gate voltages $V_{bg} = +10$~V (dotted red line), $+5$~V (dashed blue line) and $0$~V (solid black line). The pinch-off voltage ($V_{g}$ where $I_{sd} \rightarrow 0$) for Gate~1 is much less affected by the change in $V_{bg}$ indicating that this wrap-gate better screens to enclosed nanowire segment from the back-gate field, as discussed in detail in the text.}
\end{figure}

We begin by characterising the two-gate device. In Fig.~2 we plot $I_{sd}$ versus (a) $V_{g,1}$ and (b) $V_{g,2}$; in each case the other wrap-gate is grounded and the measurement is repeated at back-gate voltages $V_{bg} = +10$, $+5$ and $0$~V for reasons outlined below. Considering first the data at $V_{bg} = 0$~V (solid black lines), Gates~1 and 2 have similar pinch-off characteristics with sub-threshold slopes of $43$ and $30$~mV/dec and threshold voltages $V_{th} = -247$~mV and $-237$~mV, respectively. We use the standard approach to obtain $V_{th}$ as the voltage where a fit to the linear region of $I_{sd}$ versus $V_{bg}$ on a linear-linear graph intersects the $V_{bg}$ axis.\cite{SzeBook06} The sub-threshold slopes are only $2-3\times$ the theoretical maximum of $15.3$~mV/dec at $T~=~77$~K. The room temperature scaled sub-threshold slopes, at $117$ and $168$~mV/dec, are $50-100\%$ better than our earlier wrap-gate devices;\cite{StormNL12} this likely reflects the reduced interface charge trapping at $T = 77$~K however. The less negative $V_{th}$ that we observe here (c.f. $-670$~mV in Ref.~\cite{StormNL12}) is also consistent with reduced interface trapping.

We have also developed a method for electrically characterising the quality of the wrap-gate structures based on how the wrap-gate characteristics evolve with back-gate voltage. Nanoscale structures are rarely ideal because microscopic imperfections are at a scale comparable to the overall device structure. This is evident for the wrap-gates in Figs.~1(b-d), where holes or `mouse-bites' in the gate metallization occur due to the combined effect of granular deposition during the sputtering process and etchant attack during processing.\cite{StormNL12} The question this raises is: Are the defects just cosmetic or do they also affect performance? This motivated our method for assessing them electrically; it relies on measuring how the back-gate influences the $I_{sd}$ versus $V_{g,n}$ characteristics for a given wrap-gate segment. At $V_{g,n}~>~V_{th}$, the wrap-gated segment is strongly conducting and $V_{bg}$ has significant influence on $I_{sd}$ by gating the nanowire segments external to the wrap-gate segments.\cite{StormNL12} We see this behavior for Gates~1 and 2 in Fig.~2. In contrast, at $V_{g,n}~<~V_{th}$, the wrap-gated segment dominates the device resistance and changes in $V_{bg}$ should only have a significant effect on $I_{sd}$ if the wrap-gate segment imperfectly screens the nanowire it encloses. Imperfect screening most likely arises from holes that fully penetrate the wrap-gate metallization rather than, for example, complete removal of the underside of the wrap-gate where the device sits on the substrate. The reason why we are confident that the underside of the wrap-gate remains intact is that this is where the etchant has most restricted access to attack the metallization. Consistent with this, in our earlier work\cite{StormNL12} we saw significant beveling of the wrap-gate ends for the shortest segments (i.e., length of segment at top side of wrap-gate is much longer than on the bottom side). We observe no beveled ends in the devices we report here because the wrap-gate etch time is short in this process; as such we expect the underside of the wrap-gates to remain intact aside from holes caused by the etchant attacking the wrap-gate from the side. Indeed, now that we remove the outer oxide as the first step we expect this substrate protection to be more effective. In the earlier work,\cite{StormNL12} removal of the outer oxide after the first EBL stage could leave the wrap-gate suspended by the PMMA, enabling attack from below by the Au and W etchants. Now that we remove the entire outer oxide first, the nanowire drops down onto the substrate before any EBL, bringing the wrap-gate metal into intimate contact with the HfO2 substrate oxide, where it is better protected during the Au and W etch steps. Were the Au/W wrap-gate to lose its conformal morphology at the underside in this process, we would expect a strong back-gate effect on all wrap-gates in all devices; this is not observed in our data. 

Returning our focus specifically to the device in Fig.~1(b), one might expect both wrap-gates to show significant $V_{bg}$ dependence in the sub-threshold regime considering their appearance. However, Gate~1 is substantially less $V_{bg}$-dependent, demonstrating two important aspects regarding wrap-gates in nanowire transistors. The first is that the wrap-gate structures can tolerate significant defects yet still effectively screen the nanowire from external electric fields. The second is that the visual appearance of a wrap-gate is a poor indicator for its electrical performance. We chose the device in Fig.~1(b) specifically to highlight these points -- in this case we have a device that we might discard given its appearance, yet half of the gates work just as well as they would in a device where no wrap-gate defects are evident. We find similar behavior for our three-gate device (Fig.~1(c)); the corresponding data is shown in Supplementary Fig.~1. We suspect this defect-tolerance arises from the dual-metal recipe used. The wrap-gate metallization has a thickness of $28$~nm and while there are obvious defects in the wrap-gate surface, these may not extend through the entire metallization thickness. An interesting engineering challenge would be to model the extent to which a wrap-gate can tolerate defects, e.g., holes, oxidation, etc., without losing its effectiveness for screening the nanowire from external electric fields. As a final comment regarding the `mouse-bites' in our wrap-gate metallization; the path to minimizing them in future devices would be to improve the quality of the sputtered gate metal. We used Au here for its high conductivity and robustness against oxidation, but it tends to produce sputtered films with a large grain-size.\cite{Sputter} This in turn leads to larger voids between grains, and is likely the root cause of the `mouse-bites'. A change in the metal used for this outer layer could be beneficial; candidates would include Au-Pd alloy, Pt, Cr or Ir.\cite{Sputter} In any case, this would require finding a suitable etchant to replace the KI/\ce{I2} etch used for Au, which may be challenging. Cr and Ir are easier to wet etch, but also more susceptible to oxidation, both during sputtering\cite{Sputter} and after fabrication, reducing device lifetime. Careful optimization of the sputtering system may also enable reduced grain size.

\begin{figure}
\includegraphics[width=14cm]{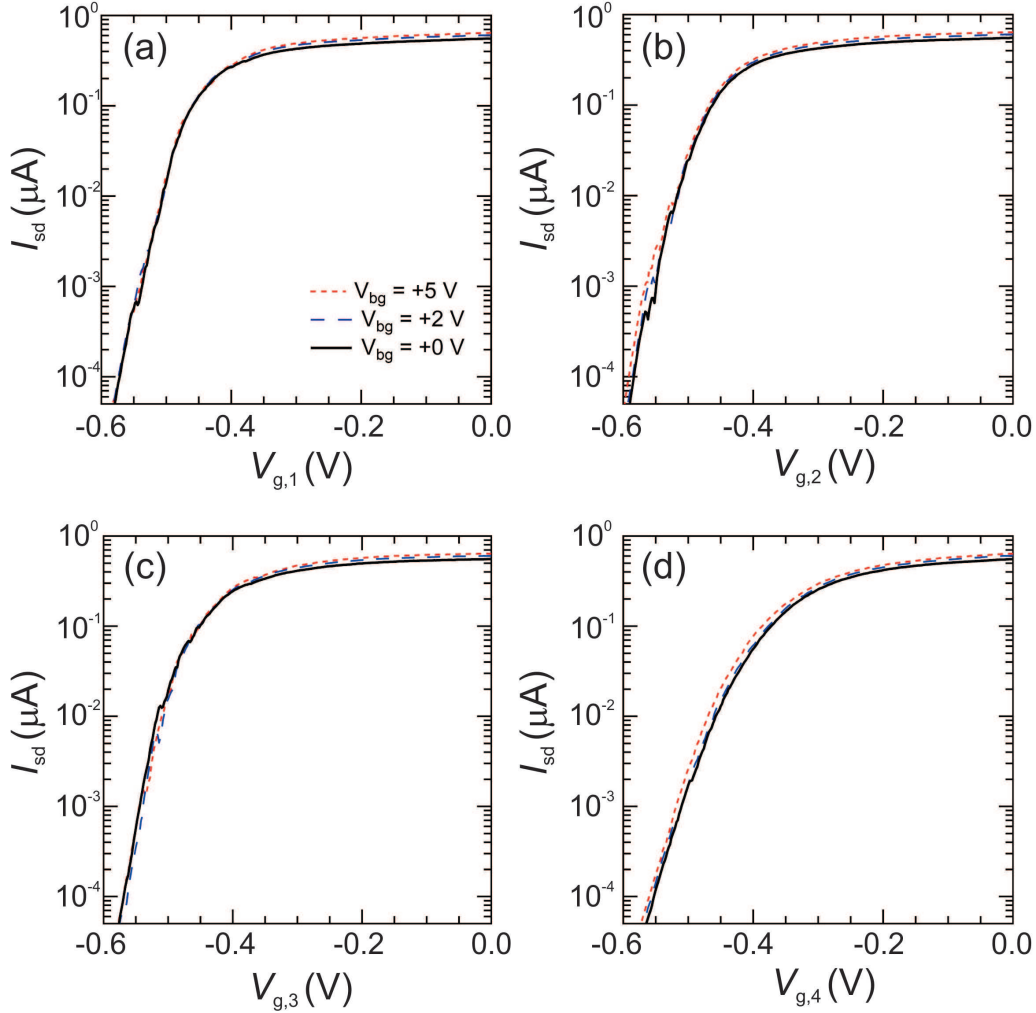}
\caption{Source-drain current $I_{sd}$ versus (a) $V_{g,1}$, (b) $V_{g,2}$, (c) $V_{g,3}$ and (c) $V_{g,4}$ at $V_{bg} = 0$~V (solid black line), $+2$~V (dashed blue line) and $+5$~V (dotted red line) for the four-gate device in Fig.~1(d). The respective threshold voltages and sub-threshold slopes are: (a) $-498$~mV, $32$~mV/dec; (b) $-501$~mV, $34$~mV/dec; (c) $-486$~mV, $25$~mV/dec; (d) $-420$~mV, $43$~mV/dec.}
\end{figure}

Figure~3 shows the individual wrap-gate characteristics for our four-gate device (Fig.~1(d)). The four gates are remarkably similar in performance, with a median threshold voltage $V_{th} = -460~\pm~40$~mV and average sub-threshold slope of $34~\pm~9$~mV/dec. Most notably, all four gates show little $V_{bg}$-dependence in the sub-threshold region, demonstrating that high-quality wrap-gates can still be achieved despite the short segment length $\sim 400$~nm and spacing between wrap-gates $\sim 200$~nm. We emphasize that the four-gate structure entails the {\it same} number of process steps as the two-gate structures, a substantial improvement on the vertical orientation where twice the number of process steps would be required.\cite{LiEDL11} The shorter gates in the four-gate device gave lower $V_{th}$ values than those in the two-gate device. This trend was observed over a range of devices; the data is presented in the Supporting Information. The maximum number of wrap-gates we have attempted is four, but with some optimization larger numbers of independent wrap-gates should be possible using the same process. The ultimate limit will be set by the ratio of the EBL defined etch resolution to the nanowire length.

While the wrap-gates effectively screen the back-gate in this device, one additional concern that arises for larger numbers of wrap-gate segments is cross-talk between adjacent back-gate segments due to the short wrap-gate segment length and relatively small separation between segments. A cross-talk experiment was conducted for the four-gate device (Fig.~1(d)) with the following methodology. Starting with Gate~1, we obtain the gate characteristic first with $V_{g,2}~=~V_{g,3}~=~ V_{g,4}~=~0$~V, and then with each of these three gates individually set to $V_{g,n}~=~-0.5$~V with the other two gates held at ground. In each of the three cases, we measure the resulting shift in threshold voltage $\Delta V_{th}$. This process is repeated for each of the three remaining wrap-gates. In the instance where cross-talk between wrap-gates is significant, we would expect the $\Delta V_{th}$ for nearest neighbours to be much greater than for next- or next-next-nearest neighbours. The extracted $\Delta V_{th}$ values range from $-30$ to $-76$~mV (full table of values in Supplementary Fig.~2) and show that the cross-talk is insignificant. The averaged shift for the nearest neighbours is $-48.7$~mV, slightly less than the average of $-59.7$~mV for the next- and next-next-nearest neighbour cells, and the opposite of what would be expected if cross-talk between adjacent gates was significant. To check whether the difference in the averages may be below certainty due to run-to-run variations, we analyzed data for $167$ sweeps obtained from $26$ different wrap-gates. We found an average $V_{th}$ variation of $4$~mV and a maximum variation of $16$~mV. The small run-to-run $V_{th}$ variation strongly supports the validity of the $\Delta V_{th}$ values found in our cross-talk analysis. Cross-talk is thus not a problem at the gate lengths and separations used in our four-gate device. However, it should become an issue at much smaller scales; we encourage theoretical work or simulations to determine the relevant length scales where this would occur.

\begin{figure}
\includegraphics[width=14cm]{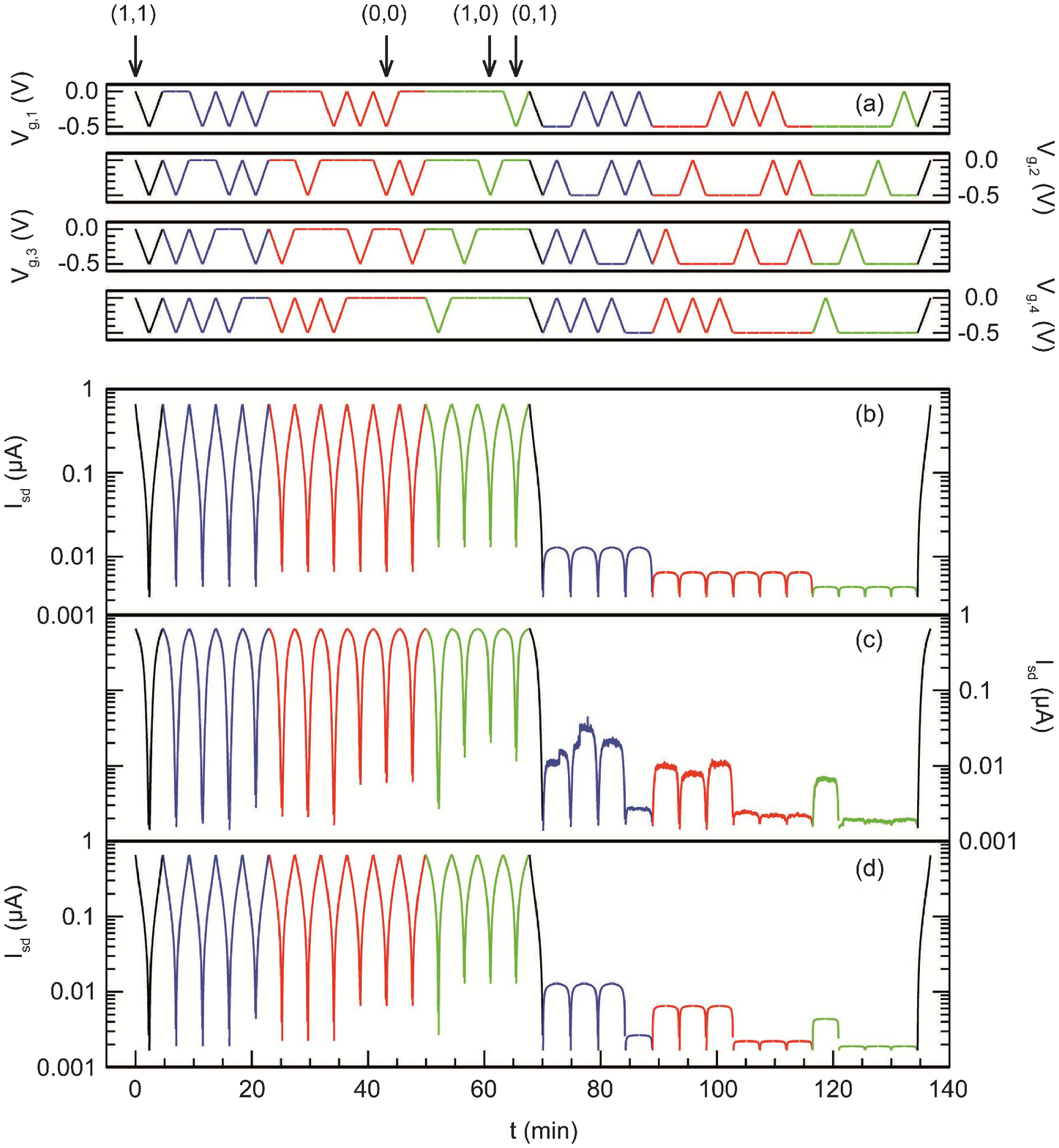}
\caption{(a) The four panels show the applied gate biases $V_{g,1}$ (top), $V_{g,2}$, $V_{g,3}$ and $V_{g,4}$ (bottom) vs time $t$. The traces are color-coded to indicate that four (black), three (blue), two (red) or one (green) wrap-gate is being swept with the remaining wrap-gates held fixed. (b) Simulated source-drain current $I_{sd}$ vs $t$ assuming the four wrap-gate segments have identical properties. (c) Measured $I_{sd}$ vs $t$ for the device in Fig.~1(d). (d) Simulated $I_{sd}$ vs $t$ assuming the segment inside Gate~4 has $1.016$ times the resistance of the other three segments (see text for details). The arrows at top indicate the operating points for the logic AND demonstration in Table~1.}
\end{figure}

\begin{table}[h]
\begin{tabular}{|c|c|c|}
\hline
Gate 1                 & Gate 2                 & Output              \\ \hline
0 ($V_{g,n} = -0.5$~V) & 0 ($V_{g,n} = -0.5$~V) & 0 ($I_{sd} = 6$~nA) \\ \hline
0 ($V_{g,n} = -0.5$~V) & 1 ($V_{g,n} = 0$~V)    & 0 ($I_{sd} = 11$~nA) \\ \hline
1 ($V_{g,n} = 0$~V)    & 0 ($V_{g,n} = -0.5$~V) & 0 ($I_{sd} = 20$~nA) \\ \hline
1 ($V_{g,n} = 0$~V)    & 1 ($V_{g,n} = 0$~V)    & 1 ($I_{sd} = 658$~nA) \\ \hline
\end{tabular}
\caption{Truth-table demonstrating logic AND operation with Gates~1 and 2 as inputs and $I_{sd}$ as output with $V_{g,3} = V_{g,4} = 0$ to `disable' these inputs for this implementation and the $0/1$ output threshold at $I_{sd} = 100$~nA. The corresponding operating points are shown by the arrows at the top of Fig.~4.}
\end{table}

To better assess the balance, temporal stability and control of these gates, we performed an in-depth time-series study of how the device behaves as various combinations of gates are swept together. The four panels in Fig.~4(a) show the voltage program for the four wrap-gates versus time $t$. Traces are color-coded to indicate whether four (black), three (blue), two (red) or one (green) wrap-gate segment was being swept with the remaining segments held fixed. When a wrap-gate segment was swept, it was taken linearly between its `on'-state ($V_{g,n}~=~0$~V) and its `off'-state ($V_{g,n}~=~-0.5$~V$~\leq~V_{th,n}$) or vice versa. Figure~4(b) shows the expected $I_{sd}$ based on a simple series resistance model consisting of five contributions: four variable contributions $R_{n}(V_{g,n})$ for the wrap-gates and a fixed contribution $R_{0}$ used to match $I_{sd}$ between experiment and simulation at $V_{g,1}~=~V_{g,2} = V_{g,3} = V_{g,4} = 0$~V (see Supporting Information for more details). For simplicity, $R_{n}(V_{g,n})$ was assumed as being linear in $V_{g,n}$. Figure~4(c) shows the measured $I_{sd}$ obtained for the voltage program in Fig.~4(a). While the general features bear strong similarity to the simple simulation in Fig.~4(b), there are differences in both the minima at lower $t$ and maxima at higher $t$ in Fig.~4(c) that point to differences in the $I_{sd}$ versus $V_{g,n}$ characteristics between the four wrap-gates. In this particular instance, Gate~4 has a higher $R_{n}(V_{g,n})$ than Gates~1-3. To demonstrate this, in Fig.~4(d) we repeat the simulation with $R_{1}(V_{g,1})~=~R_{2}(V_{g,2}) = R_{3}(V_{g,3}) = 0.984~\times~R_{4}(V_{g,4})$; this is the only linear combination of segment resistances that correctly reproduces the measured data. The imbalance is likely due to a slightly different gate capacitance for Gate~4, potentially arising from roughness in the wrap-gate metallization or slight etching of the inner wrap-gate insulator due to the outer-oxide buffered HF etch leaking through pores in the wrap-gate metallization. We expect variations between wrap-gate segments to increase as segment length is decreased or if there are imperfections in the gate metallization. Improvements in gate metallization might be obtained by moving to more openly-distributed, patterned nanowire arrays, which will limit shadowing effects whilst depositing the gate metal. A focused optimisation of the processes for deposition of the wrap-gate insulator (ALD) and gate metallization (dc sputtering), as carried out for vertical nanowire transistor arrays\cite{ThelanderTED08} might also help improve the quality and reproducibility of horizontal wrap-gate nanowire transistors. The latter would focus on reducing the grain size for the gate metallization, as discussed earlier. The data in Fig.~4 suggests it might also be possible to compensate by measuring the imbalance and scaling the off-state operating voltages $V_{g,n}$ to tune the device back into balance. We note that the data in Fig.~4(c) demonstrates significant potential for doing logic with these devices. For example, if we leave Gates~3 and 4 at ground, we can demonstrate traditional two-input logic operations using the remaining Gates~1 and 2. Table~1 presents a truth-table demonstrating a two-input logic AND gate for Gates 1 and 2 using $V_(g,n) = 0.0$~V as `1', $V_(g,n) = -0.5$~V as `0', and $I_{sd}$ as the output with $I_{sd} << 100$~nA as `0' and $I_{sd} >> 100$~nA as `1'. The four corresponding operating points are indicated by the arrows at the top of Fig.~4. Devices with more than two wrap-gate segments might provide an interesting alternative route to multi-input nanowire logic circuits,\cite{YanNat11} particularly if incorporated with the ability to make multiple contacts at different points along the sequence of wrap-gate segments (see below). Note that these devices can be operated more rapidly than in Fig.~4; we deliberately ran our device slowly here to minimise noise/hysteresis arising from the unoptimized wrap-gate structure. Vertical wrap-gate transistors featuring similar materials are capable of GHz operation with proper design/optimization.\cite{JanssonTED12}

\begin{figure}
\includegraphics[width=14cm]{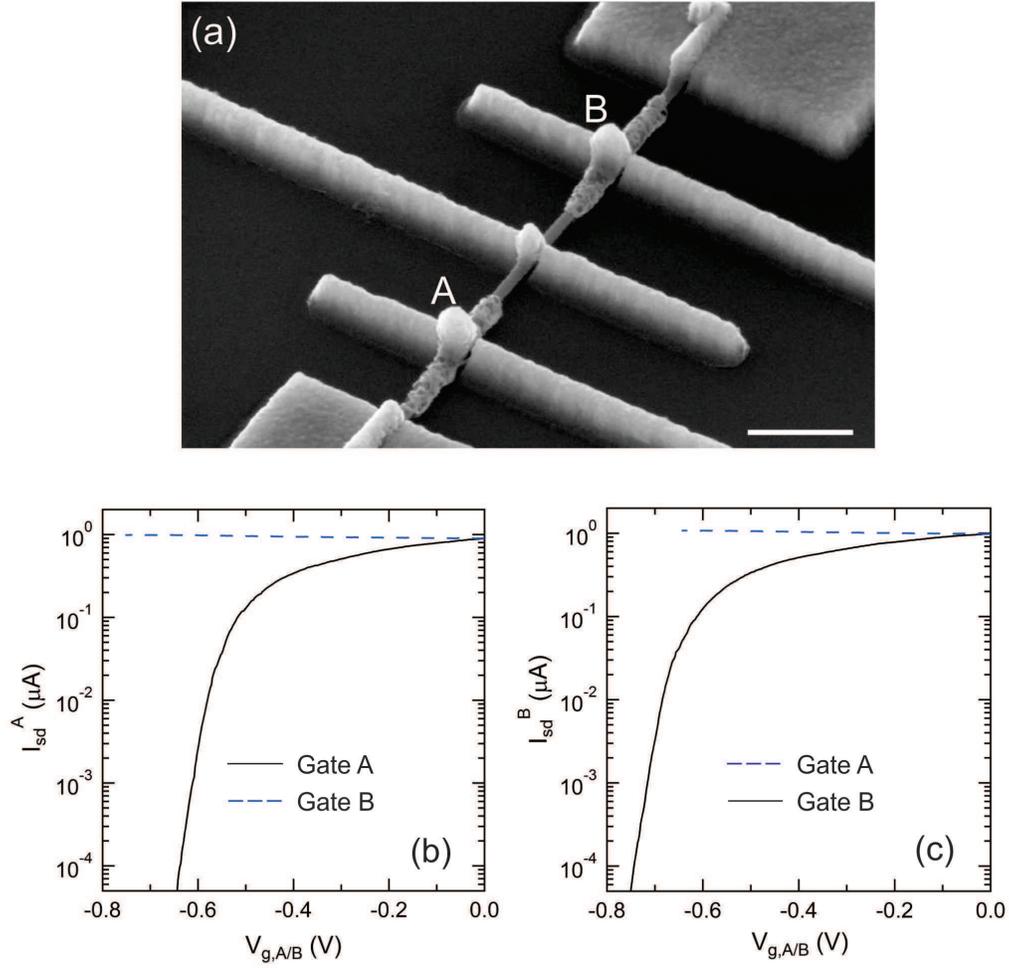}
\caption{(a) Scanning electron micrograph of a common-source wrap-gate nanowire transistor pair made using the same process used for the multiple wrap-gate structures in Fig.~1. The scale bar represents $500$~nm. (b/c) The measured source-drain current (b) $I^{A}_{sd}$ for Transistor~A and (c) $I^{B}_{sd}$ for Transistor~B versus $V_{g,A}$ and $V_{g,B}$ with $V_{bg} = 0$~V. In each case, the black solid line is for gating by a particular transistor's own gate and the blue dashed line is the response due to changes in the other transistor's gate. The bias on the unswept gate was held at zero for the data in (b/c).}
\end{figure}

Finally, in Fig.~5(a) we present a common-source wrap-gate nanowire transistor pair to demonstrate the full versatility of our fabrication method and potential for making more complex nanowire device architectures. This device was produced using exactly the same process as all three devices in Fig.~1 apart from the modification of the pattern used for the second EBL exposure to add the central source contact. The source-drain currents $I^{A}_{sd}$ through Transistor~A and $I^{B}_{sd}$ through Transistor~B were measured simultaneously, and are plotted versus the wrap-gate voltages $V_{g,A}$ and $V_{g,B}$, respectively, in Figs.~5(b/c). In each case, the solid black line demonstrates pinch-off in the transistor due to electrostatic depletion of the nanowire segment inside that transistor's wrap-gate. In contrast, if we hold a given transistor's wrap-gate fixed and sweep the other transistor's wrap-gate we see a small rise in $I_{sd}$. The bias on the unswept gate was held at zero for the data in Figs.~5(b/c). This rise is not a cross-talk effect; it occurs because the two transistors form a parallel circuit to ground. The total current flowing from the source is set by the total parallel resistance, but the share of that current flowing through Transistor~A will rise if the resistance of Transistor~B rises due to gating on that side. Note that wihout any significant effort having been devoted to optimizing their balance, the two transistors are a reasonably well matched pair: the threshold voltages differ by less than $16\%$ and the sub-threshold slopes are identical.

In conclusion, we have demonstrated a method for making horizontal wrap-gate nanowire transistors with up to four independently controllable wrap-gated segments. A key advantage of this orientation, and our approach, is that the addition of further gates does not require the addition of any extra steps to the device fabrication process. This is in stark contrast to the vertical orientation, where each additional gate involves the repetition of a significant subset of the fabrication steps.\cite{LiEDL11}. We have shown that our multiple wrap-gate transistors can be made with gates that have very similar characteristics without extensive optimization of materials or processing. There is little cross-talk between adjacent wrap-gate segments, although in some cases, imperfections in the wrap-gates mean they imperfectly screen external electric fields, e.g., those generated by a back-gate. We have also shown that the same basic process can be used to make simple multiple transistor circuits such as a common-source nanowire transistor pair. As such the method has significant potential for making more complex nanowire device/circuit architectures,\cite{YanNat11} and ultimately, towards coupling horizontal wrap-gate nanowire transistors with vertical wrap-gate nanowire transistors to achieve 3D-integrated nanowire network architectures for future electronic applications.\cite{WernerssonPIEEE10, FerrySci08}

{\bf Supporting Information.} Full details on device fabrication, data for the three-gate device in Fig.~1(c), $\Delta V_{th}$ values for the cross-talk experiment on the device in Fig.~1(d), details of the simulation model used for Figs.~4(b/d) and the relationship between $V_{th}$ and gate length are included. This material is available free of charge via the Internet at http://pubs.acs.org.

{\bf Corresponding author.} *E-mail: adam.micolich@nanoelectronics.physics.unsw.edu.au

\acknowledgement

This work was funded by the Australian Research Council (ARC), Nanometer Structure Consortium at Lund University (nmC@LU), Swedish Research Council, Swedish Energy Agency (Grant No. 38331-1) and Knut and Alice Wallenberg Foundation  (KAW). APM acknowledges an ARC Future Fellowship (FT0990285). AMB acknowledges support from the Australian Nanotechnology Network Overseas Travel Fellowship scheme. This work was performed in part using the NSW node of the Australian National Fabrication Facility (ANFF). We thank Scott Liles for assistance with the cross-talk data analysis.

\end{document}